\documentclass[twocolumn,prl,aps,showpacs,superscriptaddress,floatfix,preprintnumbers,reprint,nofootinbib]{revtex4-1}
\usepackage{color}
\usepackage{epsfig}
\usepackage{amsmath,bm}
\usepackage{blindtext}
\usepackage[modulo]{lineno}
\usepackage{xspace}

\newcommand{\bsym}[1]{\ensuremath{\boldsymbol{#1}}}
\newcommand{\ve}[1]{\bsym{#1}}
\newcommand{\epmp}{e$^{\pm}$p}
\newcommand{\bq}{\ensuremath{\bar{q}}}
\newcommand{\msbar}{\ensuremath{\overline{\mathrm MS}}}
\newcommand{\der}{\ensuremath{\mathrm{d}}}
\newcommand{\sub}[2]{\ensuremath{#1_{\mathrm{#2}}}}
\newcommand{\Sup}[2]{\ensuremath{#1^{\mathrm{#2}}}}

\newcommand{\credib}[2]{\ensuremath{\{#1\text{--}#2\}}}

\newcommand{\newpar}{\newline\indent}

\newcommand{\mbstrut}[1]{\rule{0mm}{#1}}

      \makeatletter
      \newcommand*{\textalltt}{}
      \DeclareRobustCommand*{\textalltt}{%
	      \begingroup
	      \let\do\@makeother
	      \dospecials
	      \catcode`\\=\z@
	      \catcode`\{=\@ne
	      \catcode`\}=\tw@
	      \verbatim@font\@noligs
	      \@vobeyspaces
	      \frenchspacing
	      \@textalltt
      }
      \newcommand*{\@textalltt}[1]{%
	      #1%
	      \endgroup
      }
      \makeatother

\newcommand{\Figref}[1]{Fig.~\ref{#1}}

\newcommand{\Tabref}[1]{Table~\ref{#1}}

\nolinenumbers

\begin{document}

\title{New constraints on the up-quark valence distribution in the proton}
\author{Ritu Aggarwal}\email{ritu.aggarwal1@gmail.com}
\affiliation{USAR, Guru Gobind Singh Indraprastha University, Delhi, India}

\author{Michiel Botje}\email{m.botje@nikhef.nl}
\affiliation{Nikhef, Amsterdam, The Netherlands}

\author{Allen Caldwell}\email{caldwell@mpp.mpg.de}
\affiliation{Max-Planck-Institut f\"ur Physik, M\"unchen, Germany}

\author{Francesca Capel}\email{capel@mpp.mpg.de}
\affiliation{Max-Planck-Institut f\"ur Physik, M\"unchen, Germany}

\author{Oliver Schulz}\email{oschulz@mpp.mpg.de}
\affiliation{Max-Planck-Institut f\"ur Physik, M\"unchen, Germany}

\begin{abstract}

The high-$x$ data from the ZEUS Collaboration 
are used to extract parton density distributions of the proton deep in the perturbative regime of QCD.  The data primarily constrain the up-quark valence 
distribution and new results are presented on its $x$-dependence as well as on the momentum carried by the up-quark.
The results were obtained using 
Bayesian analysis methods which can serve as a model for future parton density extractions.
\end{abstract}

\date{\today}
\maketitle

\section{Introduction}
The inner structure of hadrons continues to be a subject of intense interest, with many teams pursuing quantitative studies~\cite{MMHT:1,MMHT:2,CT:1,CT:2,ABM:1,ABM:3,NNPDF:1,NNPDF:2,herapdf2.0:2,herapdf2.0}. Although theoretically motivated expectations exist for some aspects of the structure, 
precise theoretical calculations remain challenging (see, e.g., \cite{Constantinou:2020hdm}) so that
the quantitative behavior is effectively determined from data.
The large-$x$ behavior of parton distributions in the proton has received increased
interest in recent years~\cite{ref:Balletal,Courtoy:2020fex}
and deviations from the expectations from the spectator counting rules of
Brodsky and Farrar~\cite{BF} are a focus of attention.
 
Our work is based  on the high-$x$ \epmp\ scattering data from the ZEUS
Collaboration~\cite{ref:ZEUS1,ref:ZEUS2} which is unique in providing measurements up to $x=1$ in the high $Q^2$ regime.
The data are in a kinematic region where the perturbative QCD evolution equations~\cite{Gribov:1972ri,Gribov:1972rt,Lipatov:1974qm,Dokshitzer:1977sg,
Altarelli:1977zs} are valid so that a conceptually clean analysis can be performed.
Our results can furthermore serve as inputs for more global
analyses~\cite{PDFsummary:1,PDFsummary:2,PDFsummary:3}.
Novel techniques were developed to carry out the present analysis~\cite{ref:PartonDensity}. 

The \epmp\ cross section as $x \rightarrow 1$ is dominantly 
from scattering off the two up and one down valence quarks.
Here the contribution from the up quarks is about 
one order of magnitude larger than that from down-quarks because 
the up quark electric charge is twice that of the down quark.
We therefore focus on results on the valence up-quark distribution
and report, in particular, on its power-law behavior as 
$x\rightarrow 1$.  We also extract the total momentum carried by the 
valence up- and down-quark distributions, as well as that of the
gluon distribution.  These results are compared to the well-known expectations that,
at asymptotically high energies, the gluons and quarks should carry approximately the
same momentum (see, e.g., \cite{HM}).  

The kinematics of inclusive \epmp\ scattering events are defined from
the four-momenta $p$ of the incoming proton and $k$ ($k'$) of the incoming (scattered) lepton. From these 
are computed the scaling variables
$Q^2 = -q^2 = -(k-k')^2$ and $x =  Q^2/(2p\cdot q)$, and events are typically
accumulated in bins of $x$ and $Q^2$.
The ZEUS data used in this analysis cover the range $0.03 \leq x \leq 1$ and 
$650 \leq Q^2 \leq 20000$~GeV$^2$.

Scattering at large values of $Q^2$ 
probes the proton at short (transverse) distance scales, with $Q=1$~GeV corresponding approximately to a scale of $0.2$~fm.  
In a frame where the proton has very large momentum
the Bjorken-$x$ variable has an intuitive interpretation  
as the fractional momentum carried by the struck parton (quark $q$, antiquark
$\bar{q}$ or gluon $g$)
in the scattering process.

\section{Analysis Procedure}

The analysis was done in a forward modeling approach where the
\epmp\ cross-sections are computed from the parton distributions and,
after correction for radiative and detector effects, used to predict the event
numbers in the experimental set of \mbox{$x$-$Q^2$} bins. These predictions are
then compared to the measured 
number of events in a Bayesian analysis of the data.  

To compute the cross-sections,
we parameterize a set of quark, antiquark and gluon distributions 
$xf_i(x)$ at a fixed 
value of $Q^2_0 = 100$~GeV$^2$, with free parameters
to be determined from the data. 
Here $f_i(x)$ is the number density of partons
of type $i$, and $xf_i(x)$ is the fractional momentum density of these partons.

The QCDNUM~\cite{ref:QCDNUM} program is used to evolve these distributions
to larger values
of $Q^2$ in the \msbar-scheme at NNLO in perturbative QCD~\cite{QCD:1,QCD:2,
QCD:3,QCD:4,QCD:5,QCD:6,QCD:7,QCD:8,QCD:9,QCD:10}.
Because the data, and $Q^2_0$, are well above (below) bottom (top) quark mass
thresholds, the
evolution is carried out with $5$ active quark flavors $i = (u,d,s,c,b)$. 

In the analysis we impose the momentum
sum rule and the valence quark
counting rules. The momentum sum rule states that the fractional 
momenta of all partons in the proton add-up to unity:
\begin{equation}
   \sum_i \int_0^1  xf_i(x) dx =  \sum_i \Delta_i = 1 \; .
\end{equation}
We introduce here the notation $\Delta_i$ for the total momentum fraction carried by
the  parton species~$i$.

The quark counting rules fix the net number of quarks in the proton so
that its quantum numbers are conserved:
\begin{equation}\label{eq:vsum}
  \int_0^1 [ q_i(x) - \bar{q}_i(x) ]\; \der x = \left\{
  \begin{array}{l}
    2\ \ \text{for}\ \ i = u\; , \\
    1\ \ \text{for}\ \ i = d\; ,\\
    0\ \ \text{for}\ \ i = s,c,b\; .
  \end{array}
  \right.
\end{equation}
Here and in the following we use the notation $q$, $\bar{q}$ and $g$
to denote quark, antiquark and gluon densities.
 
It is important to point out that the QCD evolution equations
guarantee that sum rules which are imposed
at the starting scale $Q^2_0$ will be satisfied at
all scales. 

\section{Parametrizations}

To parameterize the quark densities it is convenient to write them
as valence (\Sup{q}{v}) and sea ($\Sup{q}{s}$) distributions,
\[
  q + \bar{q} = (q - \bar{q}) + 2\bar{q} = \Sup{q}{v} + \Sup{q}{s}.
\] 
We parameterize the valence momentum distributions as
\begin{equation}\label{eq:xvalence}
  x\Sup{q}{v}_i(x,Q^2_0) = \left\{ \begin{array}{ll}
     A_i \;x^{\lambda_i} (1-x)^{K_i} & \ \text{for} \ i = u,d \\
     0                               & \ \text{otherwise},
     \end{array}\right.
\end{equation}
where $A_i$, $\lambda_i$ and $K_i$ are to be determined from the data.
Replacing $\lambda_i$ by $\lambda_i - 1$ above yields parametrizations
for the $\sub{u}{v}$ and $\sub{d}{v}$ number densities, needed to compute the
valence sum rules. 
The integral of this number density is finite for $\lambda_i > 0$. 
 
We choose a similar parametrization for the antiquark distributions:
\begin{equation}\label{eq:xqbar}
    x\bar{q}_i(x,Q^2_0) = A_i\; x^{\lambda_{\bar{q}}} (1-x)^{K_{\bar{q}}}
    \quad \text{for} \quad i = \bar{u},\bar{d},\bar{s},\bar{c},\bar{b},
\end{equation}
where all antiquark flavors have the same $x$-dependence but different
normalizations.   This is acceptable since we are fitting data in a limited
large-$x$ range
where there is no
sensitivity to the different quark flavors. We must have $-1 < \lambda_{\bar{q}} < 0$ for $x\bar{q}$ to be integrable and increasing at low $x$.

We parameterize  the gluon density as the sum of a valence and sea component, 
with the valence (sea) gluon dominating at large (small) values of $x$:
\begin{eqnarray}
    xg(x,Q^2_0) &=& x\Sup{g}{v}(x) + x\Sup{g}{s}(x) \nonumber \\
    &=& \Sup{A}{v}_g\; x^{\Sup{\lambda}{v}_g}(1-x)^{K_{g}}
    + \Sup{A}{s}_g\; x^{\Sup{\lambda}{s}_g}(1-x)^{K_{\bar{q}}}\quad
\end{eqnarray}
To obtain an integrable gluon density with a valence (sea) term that
decreases (increases) towards low $x$ we set
\mbox{$\Sup{\lambda}{v}_g > 0$} and $ -1 < \Sup{\lambda}{s}_g < 0$.
For the valence component, we keep the
power $K_g$ of $(1-x)$ as a free parameter, while the $(1-x)$ power
of the sea component is set to the $K_{\bar{q}}$ of the antiquark densities.

This choice of parametrizations yields densities that
are positive-definite. Furthermore, the sum rules are easy 
and fast to evaluate as Beta functions, and are used to restrict the number 
of free parameters as follows.

The valence sum rule (\ref{eq:vsum}) relates the valence normalizations to
the shape parameters by
\begin{equation}
A_i = \Sup{N}{v}_i\; \frac{\Gamma(\lambda_i+K_i+1)}{\Gamma(\lambda_i)\Gamma(K_i+1)}
\qquad i = u,d\;,
\end{equation}
with $\Sup{N}{v}_u = 2$ and $\Sup{N}{v}_d = 1$.

Integrating the valence momentum densities and using the property
$\Gamma(z+1) = z\Gamma(z)$ we find the following relation between
the total momentum $\Delta$ carried by the $u$ or $d$ valence quarks
and the shape parameters:
\begin{equation}\label{eq:getlambda} 
\Delta_i =  
\Sup{N}{v}_i\; \frac{\lambda_i}{\lambda_i + K_i + 1} \qquad i =u,d\
\end{equation}
which allows us to fix the $\lambda_i$ by fitting $\Delta_i$ and $K_i$.

Likewise we can fix the normalizations of the antiquark and gluon densities
by fitting their momentum fractions $\Delta$, subject to the momentum
sum rule constraint
\begin{equation}\label{eq:deltasum}
  \Delta_u + \Delta_d + 2 \sum_{\bar{q}}
  \Delta_{\bar{q}} + \Sup{\Delta}{v}_g + \Sup{\Delta}{s}_g = 1, 
\end{equation}
where the sum runs over $\bar{q} = \{\bar{u},\bar{d},\bar{s},\bar{c},\bar{b}\}$.
The 9 momentum fractions and 7 shape parameters 
\[
  K_u,\ K_d,\ \lambda_q ,\ \Sup{\lambda}{v}_g,\ \Sup{\lambda}{s}_g,\ K_{\bar{q}},\ \text{and}\  K_g
\]
give in total 16 parameters to be fitted to the data.  Together with the constraint (\ref{eq:deltasum}) this corresponds to 15 degrees of freedom in the fit.

\section{Predicted number of events}

As mentioned above, the parton densities at $Q^2_0 = 100$~GeV$^2$ 
were evolved upward at NNLO
in the fixed 5-flavor scheme with the program QCDNUM. 
The strong coupling constant was set to $\sub{\alpha}{s} = 0.118$ at $M_Z^2$.

From the evolved densities the $F_2$, $F_{\mathrm{L}}$ and
$xF_3$ structure functions were evaluated at NNLO~\cite{F123:1,
F123:2,F123:3,F123:4,F123:5,F123:6} and used to compute the neutral current \epmp\ cross-sections at the
Born level. Multiplying by the luminosity of the ZEUS data sets then gives a vector \ve{\nu} of binned event expectations.

As described in~\cite{ref:ZEUS2} by the ZEUS collaboration, the vector \ve{n}
of expected number of events is calculated from 
\[
  \ve{n} = \ve{R}\ve{T}\; \ve{\nu} 
\]
where \ve{R} and \ve{T} are transfer matrices that account for radiative and
detector/reconstruction effects, respectively, and map the binning at the Born level to the coarser binning at the observed level.  
This procedure yields the expected number of events in the 153 bins defined
by ZEUS for each data set.

The systematic uncertainties in the ZEUS event reconstruction are
given in~\cite{ref:ZEUS2} as deviations from the nominal transfer matrix \ve{T}.
The total deviation is then a sum of 10 deviation matrices, weighted by
a set of 10 systematic parameters $\delta_i$ that are left free in the fit.

\section{Parameter Extraction}
Assuming that the counts in a bin $i$ are Poisson distributed with a mean equal
to the expected number $n_i$ of events, we can compute the probability of observing
the actual data, given the values of the parameters.
Using Bayes' theorem we then calculate the joint posterior probability density of the parameters, given the data,
with as input the prior probabilities of the parameters. 
Single-parameter distributions or correlations among the parameters are evaluated by integrating over the other parameters. 
The posterior is not only conditional on the data, but also on all the assumptions made in the analysis, such as the choice of parametrizations.

Sound prior knowledge and known physical constraints are easily implemented in the Bayesian approach, but priors should be chosen with care to not introduce unwanted effects in the posterior probability density. 
Comparing the posterior and prior probability densities provides an
easy way to judge
the information content of the data.%

A 9-dimensional Dirichlet distribution~\cite{ref:Betancourt} with 9 shape
parameters $\alpha_i$
was chosen for the prior of the momentum components $\Delta_i$. 
Note that a Dirichlet distribution ${\rm Dir}(\ve{\alpha})$ of $k$ 
independent variables $x_i \in [0,1]$  
lives on a $(k-1)$-dimensional manifold defined by
$\sum x_i = 1$.\footnote{
  A Dirichlet distribution is a multivariate generalization of the Beta distribution.
  For instance ${\rm Beta}(\alpha_1,\alpha_2)$ of one variable $x$ is the same as 
  ${\rm  Dir}(\alpha_1,\alpha_2)$ of two variables
  $(x_1,x_2)$ with $x_1 + x_2 = 1$.
}
With a Dirichlet prior
the sum rule (\ref{eq:deltasum}) is thus automatically satisfied.
The choice of the parameters $\alpha_i$ was guided by the expectation that,
asymptotically, gluons and quarks should carry approximately
the same momentum, that valence up quarks should carry about twice
the momentum of valence down quarks, and that the heavier quarks should carry
little momentum.

The parameters $\alpha_i$ of the Dirichlet distribution are given in 
Table~\ref{tab:priors}, 
%
\begin{table}[b]
\caption{Priors used in the parton density fit for all parameters. There are 9
parameters in the vector \ve{\alpha} and 10 in \ve{\delta}. 
The normal distributions are truncated to the
range indicated and their mean and sigma are given in brackets.}
\begin{ruledtabular}
\begin{tabular}{cll}
   & \textbf{Prior}  & \textbf{Range}\\
 \hline
$\ve{\alpha}$        & Dir(20,\;10,\;20,\;20,\;5,\;2.5,\;1.5,\;1.5,\;0.5) & $[0,1]$       \\
$K_u$                & Normal(3.5, 0.5)                             & $[2,5]$       \\
$K_d$                & Normal(3.5, 0.5)                             & $[2,5]$       \\
$\Sup{\lambda}{v}_g$ & Uniform                                      & $[0,1]$       \\
$\Sup{\lambda}{s}_g$ & Uniform                                      & $[-1,-0.1]$   \\
$K_g$                & Normal(4, 1.5)                               & $[2,7]$       \\
$\lambda_{\bar{q}}$          & Uniform & $[-1,-0.1]$ \\
$K_{\bar{q}}$                & Normal(4, 1.5) & $[3,10]$     \\
$\ve{\delta}$         & Normal(0, 1)                                 & $[-5,5]$      \\
\end{tabular}
\end{ruledtabular}
\label{tab:priors}
\end{table}%
%
together with the prior distributions of all other
parameters. Also listed in the table are the parameter ranges imposed. Note that
the $\lambda$ ranges are set such that all parton distributions are integrable
and either vanish, or increase at low $x$, as required.

The parameters $K_u$ and $K_d$ determine the behavior of the valence quark
distributions as $x \rightarrow 1$, and their values are of great interest.
As given in \Tabref{tab:priors},
we choose as prior a truncated Normal distribution that accommodates
a range of about $K = 3$ to 4 at $Q^2=10$~GeV$^2$, as  
found by different global fitting groups~\cite{ref:Balletal}.
At our $Q^2$ we expect a somewhat larger value for $K$ than at 10~GeV$^2$.
The Brodsky-Farrar counting rules predict that $K_{\bar{q}} \approx K_g+1 \approx K_{u,d}+2$.  Our priors are set to accomodate this expectation.
Note that all our prior choices for the $K$-parameters fulfill the requirement that
the parton densities, 
as well as their first derivatives, go to zero as 
$x\rightarrow 1$~~\cite{ref:Balletal}. 

The extraction of the parameters of the parton distributions was performed using the Bayesian Analysis Toolkit (BAT.jl)~\cite{Schulz:2021BAT}. 
A Markov Chain Monte Carlo technique was used to sample the posterior probability distribution in the space of the parameters.  The accuracy of this data analysis setup was validated using simulated data sets.

\begin{figure*}[htb] 
    \centering
    \includegraphics[width=0.9\textwidth]{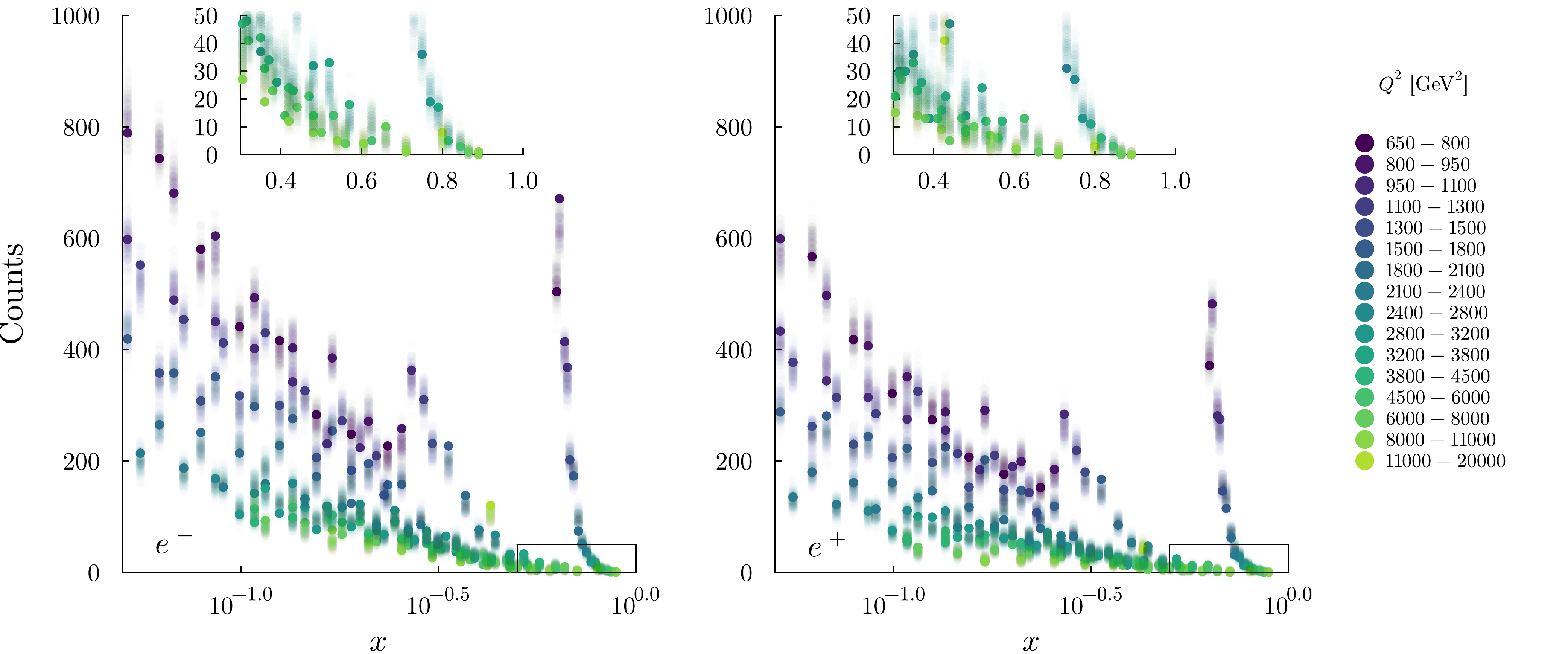}
    \caption{Predictions of event numbers calculated from sampling the posterior
         probability distribution of the fitted parameters (shaded bands) compared
         to the ZEUS data (filled dots) for e$^-$p (left) and e$^+$p scattering
         (right). The predictions and data are displayed at the center of the bins in which events are recorded.
         For clarity the boxes at large $x$ are shown enlarged in the insets.}
    \label{fig:dataspace}
\end{figure*}

A comparison of the measured data to event numbers predicted from the posterior probability distribution is shown
in \Figref{fig:dataspace}. The predicted counts from the model cover the observed counts well.  We evaluated Pearson's $\chi^2$ for the two data sets using the global mode parameters to evaluate the predicted number of counts, with resulting values of $\chi^2_P=322$ from fitting the 306 event numbers.   Tests with simulated data yield a $p$-value of $0.23$, indicating good agreement between the model and the data.
%

\section{Results}

As discussed in the introduction, the ZEUS high-$x$ data primarily set constrains on the valence distribution of the up quark, and we focus on the relevant parameters here.  A summary of the full set of results is given in \Tabref{tab:results}. 

\begin{table}[b]
\caption{Parameter values obtained from this analysis. For each parameter
is given the value of the mode of the joint posterior and
of its marginal distribution, with errors corresponding to the
68\% smallest credible interval. The fit does not constrain the values of
$\Delta_{\bar{s},\bar{c},\bar{b}}$,
$\Sup{\lambda}{v}_g$ and $\Sup{\lambda}{s}_g$ (see text).}
\begin{ruledtabular}
\begin{tabular}{lcclcc}
  & \textbf{Global}& \textbf{Marginal} & 
  & \textbf{Global}& \textbf{Marginal} \\  
  & \textbf{mode} & \textbf{mode} & & \textbf{mode} & \textbf{mode} \\
 \hline
$\Delta_u         $ & 0.225  & $ 0.219_{-0.009}^{+0.009}$ &\quad \mbstrut{1.2em}
$K_u              $ & 3.89   & $ 3.74_{-0.13}^{+0.18}   $ \\
$\Delta_d         $ & 0.084  & $ 0.092_{-0.026}^{+0.023}$ &\quad \mbstrut{1.5em}
$K_d              $ & 3.18   & $ 3.51_{-0.42}^{+0.53}   $ \\
$\lambda_{\bar{q}}$ & $-$0.50& $-0.54_{-0.09}^{+0.09}   $ &\quad \mbstrut{1.5em}
$K_{\bar{q}}      $ & 7.42   & $ 6.38_{-1.42}^{+1.17}   $ \\
$K_{g}            $ & 4.69   & $ 5.02_{-1.21}^{+1.21}   $  & & & \mbstrut{1.5em} \\
$2\Delta_{\bar{u}}$ & 0.092  & $ 0.100_{-0.024}^{+0.026}$ &\quad \mbstrut{1.5em}
$2\Delta_{\bar{d}}$ & 0.032 & $ 0.022_{-0.014}^{+0.022}$ \\
$\Sup{\Delta}{v}_g$ & 0.250 & $ 0.245_{-0.044}^{+0.040}$ &\quad \mbstrut{1.5em}
$\Sup{\Delta}{s}_g$ & 0.275  & $ 0.251_{-0.045}^{+0.040}$ \\
\end{tabular}
\end{ruledtabular}
\label{tab:results}
\end{table}%
%
We remind the reader that the parton densities are defined in the \msbar\ scheme at
NNLO and parameterized at a scale of $Q^2_0 = 100$~GeV$^2$.

Not listed in \Tabref{tab:results} are
the parameters
$\Delta_{\bar{s},\bar{c},\bar{b}}$,
$\Sup{\lambda}{v}_g$ and $\Sup{\lambda}{s}_g$ 
since no significant reduction of the prior range was observed in the posterior. We find
68~\% probability upper limits $2\Delta_{\bar{s}} < 0.027$, 
$2\Delta_{\bar{c}} < 0.038$ and $2\Delta_{\bar{b}} < 0.007$.
The prior ranges of
$\Sup{\lambda}{v}_g$ and $\Sup{\lambda}{s}_g$ are given in~\Tabref{tab:priors}. 

\begin{figure}[b]
    \centering
    \includegraphics[width=\columnwidth]{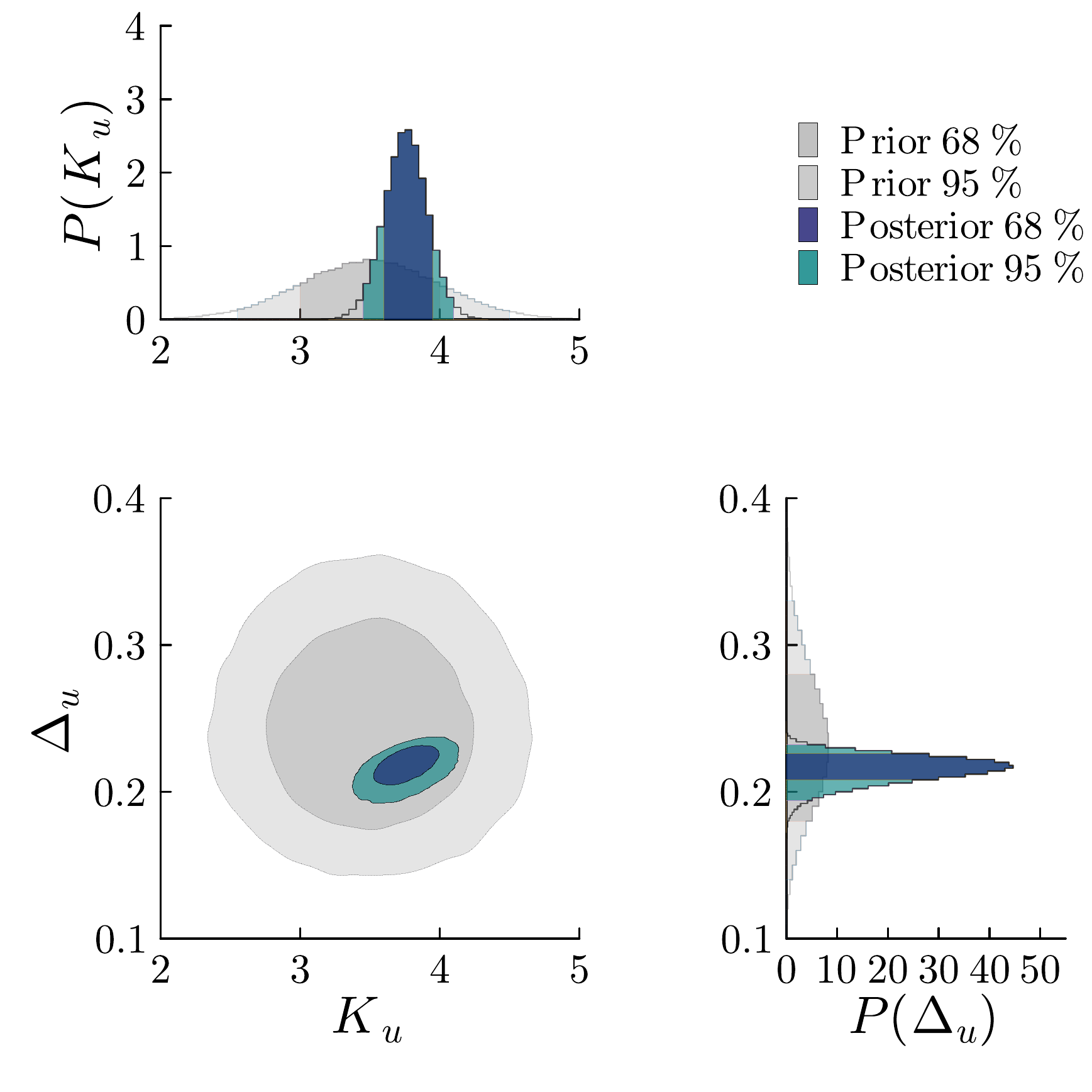}
    \caption{
     Joint probability contours of the up-valence parameters  
     $K_u$ and $\Delta_u$ (lower left) and the marginal distributions of
     $K_u$ (top) and $\Delta_u$ (lower right). Shown are the 68 and 95\% smallest
     credible intervals for both the prior (light shaded) and
     posterior (dark shaded) distributions.
    }
    \label{fig:momenta}
\end{figure}

Figure~\ref{fig:momenta}
%
%
shows a comparison of the prior and posterior probability contours for $K_u$ and $\Delta_u$ as well as their marginal distributions.  As is clear in the figure, the data strongly constrains these parameters. 
The momentum fraction carried by the valence up quark distribution was found to be
$\Delta_u \approx 0.22$ with a credible interval 
of \credib{0.210}{0.228} where, unless otherwise stated, we will in the following 
always refer to the smallest intervals with 68\% probability content.
Such a high precision measurement of~$\Delta_u$ has, to the best of our knowledge,
not yet been reported in the literature.

We obtained for 
the power of the up-valence $(1-x)$ component a value of $K_u \approx 3.8$ with a credible interval of \credib{3.61}{3.92}.
We subdivided the data into a set of bins containing data with $x\geq 0.5$ and the complement. The resulting intervals were \credib{3.40}{4.03}(\credib{3.71}{4.52}) for the lower- (higher-)$x$ data, indicating that the higher-$x$ data indeed provide valuable new information.
\newpar
The summary of previously measured $K_u$ values in~\cite{ref:Balletal} includes results
with stricter bounds than reported here. However, these were determined from data
at much lower $Q^2$ where higher-twist effects may play a role, and which do not
extend to the highest values of $x$, as do the present data.
Furthermore in~\cite{ref:ZEUS2} it is shown that different \mbox{high-$x$} parton
distributions do not overlap within their quoted uncertainties. 
\begin{figure}[tb]
    \centering
    \includegraphics[width=\columnwidth]{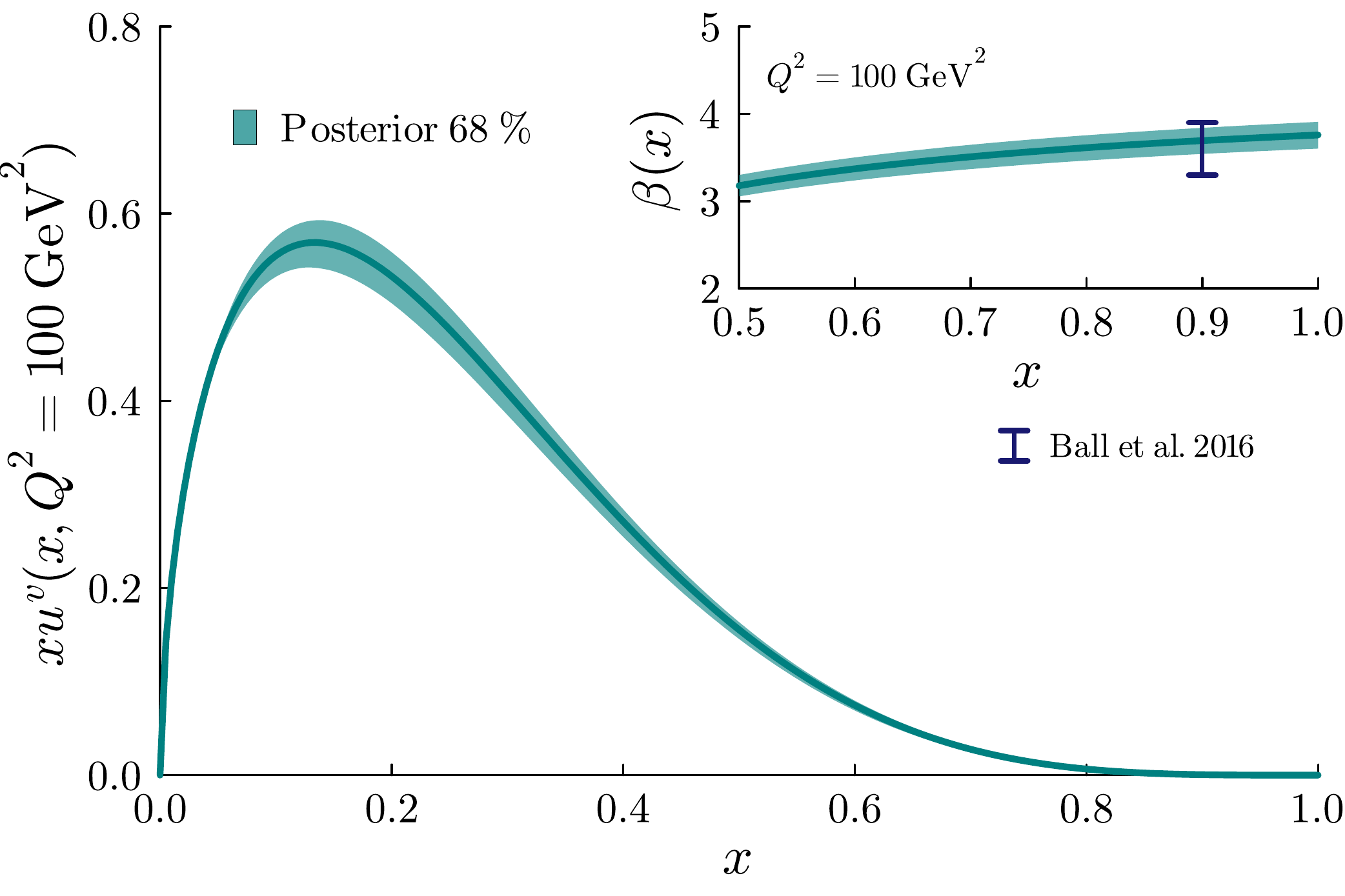}
    \caption{The valence distribution $x\Sup{u}{v}$ and the effective
    $(1-x)$ power $\beta$ from this analysis at a value 
    of $Q^2 = 100$~GeV$^2$. The 68\% smallest probability contour is
 depicted. 
    The vertical range 
    plotted in the inset
    summarizes the results on $\beta$ reported in \cite{ref:Balletal}.
}
    \label{fig:EffExp}
\end{figure}
%
\newpar
We follow the suggestion given in~\cite{ref:Balletal} and calculate the effective power of $(1-x)$:
\begin{equation}
    \beta(x,Q^2_0)=\frac{\partial \ln [x\Sup{u}{v}(x,Q^2_0)]}
    {\partial \ln(1-x)}
\end{equation}
%
and show the result in Fig.~\ref{fig:EffExp}
where we also compare our results to a summary of 
$\beta$-values at $x=0.9$
reported in ~\cite{ref:Balletal}. 
Our analysis agrees well with
these results, and provides a more constrained range of values. 
It should be noted, however, that our result for $\beta(x)$
is not only constrained by the data, but 
possibly also by the form of the parametrization chosen and that such a parametrization dependence is left for investigation in a future analysis.

 Figure~\ref{fig:EffExp} also shows our result on the
$x\Sup{u}{v}$ distribution. Although plotted down to a value of $x = 0$, 
we remind the reader
that our parametrizations are not meant to describe the parton densities
at very small values of $x$, as we have only analyzed data in the range $x > 0.03$.

Using (\ref{eq:getlambda}) to compute  $\lambda_u$ from $\Delta_u$ and $K_u$ 
gives a value of $\lambda_u \approx 0.58$ with 
a credible interval of \credib{0.53}{0.62}.
\newpar
For the
momentum fractions besides that of the valence up-quark, we obtain
\[
\begin{array}{rll}
  \Delta_d          & \approx 0.11 &\qquad \credib{0.07}{0.12}\\
  \Delta_{\rm sea} = 2 \sum_{\bq} \Delta_{\bq} & 
             \approx 0.17 & \qquad \credib{0.16}{0.22}  \mbstrut{1.2em} \\
  \Delta_g = \Sup{\Delta}{v}_g + \Sup{\Delta}{s}_g & 
             \approx 0.50 &\qquad \credib{0.47}{0.53}. \mbstrut{1.2em}
\end{array}
\]  
With the fitted data, there is little sensitivity to the gluon density and the approximately equal momentum sharing between the quarks and gluons is largely a consequence of our prior choice on $\Delta_g$.

The $(1-x)$ powers for the valence down-quark and the gluon were found to be 
$K_d\approx 3.5$ and $K_g\approx 5.0$ with credible intervals \credib{3.1}{4.0} and
\credib{3.8}{6.2}, respectively.   The sea densities had a power of $(1-x)$ of $K_{\bar{q}} \approx 6.4$ with credible interval \credib{5.0}{7.6}.
These values are in line with the expectation
that $K_d \approx K_u$ and that $K_g>K_{u,d}$ and $K_{\bar{q}} > K_g$.

\section{Summary}

We have performed the first analysis of the ZEUS high-$x$ data set~\cite{ref:ZEUS1}
and extracted precise information on the momentum content and $x$ dependence of the
valence up-quark distribution at the highest values of $x$.
The analysis was based on a forward modeling approach, 
taking parton distributions at
a scale of $Q^2_0=100$~GeV$^2$ and evolving them upward 
with the QCDNUM program at NNLO in QCD perturbation
theory.  The evolved parton distributions were then used to predict expected event
numbers in the measurement intervals used by the ZEUS Collaboration, and a Poisson
probability was evaluated for the individual measurements.  The full posterior probability distribution of the parameters of parton distributions was then evaluated
in a Bayesian fit using the BAT.jl package.  All systematic uncertainties associated with the data were included in this analysis.
\newpar
The high-$x$ data from ZEUS primarily inform us on the valence up-quark distribution.
Given our
simple parametrizations, which are adequate for the data analyzed, we
obtain a precision of 
$1$~\% on the total momentum carried by the
up-valence quarks, while the
power of $(1-x)$ was found to be in the range 
$3.6-3.9$ at $Q^2=100$~GeV$^2$.  
These results, based on data at the highest values of $x$, represent a significant
step in our understanding of the proton.

\vspace{-0.3cm}

\section{Acknowledgments}

\vspace{-0.3cm}

The authors thank Andrii Verbytskyi for his help in the technical developments of the PartonDensity.jl package, as well as Amanda Cooper-Sarkar, Pavel Nadolsky and Aurore Courtoy for informative discussions. We are especially grateful to I.~Abt for her help in identifying an error in our data set selection in an earlier version of this manuscript.  We thank T. Rogers for his comment on the positivity requirement of \msbar\ parton densities.
R. Aggarwal acknowledges the support of Savitribai Phule Pune University. F. Capel was employed by the Excellence Cluster ORIGINS, which is funded by the Deutsche Forschungsgemeinschaft (DFG, German Research Foundation) under Germany's Excellence Strategy - EXC-2094-390783311, during much of the project duration.

\bibliography{literatur.bib}
\end{document}